\documentclass[aps,prl,superscriptaddress,twocolumn,showpacs]{revtex4}

\usepackage{amsmath,amsfonts,bm,graphicx}

\newcommand{\eps}{\epsilon}
\newcommand{\om}{\omega}

\begin{document}

\title{Quantum Shuttle in Phase Space}

\author{Tom\'a\v s Novotn\'y}
\email{tno@mic.dtu.dk}

\affiliation{Mikroelektronik Centret,
         Technical University of Denmark,
         \O rsteds Plads, Bldg.\ 345 east,
         DK-2800 Kgs.\ Lyngby, Denmark}

\affiliation{Department of Electronic Structures,
     Faculty of Mathematics and Physics, Charles University,
     Ke Karlovu 5, 121 16 Prague, Czech Republic}

\author{Andrea Donarini}
\email{ad@mic.dtu.dk}

\affiliation{Mikroelektronik Centret,
         Technical University of Denmark,
         \O rsteds Plads, Bldg.\ 345 east,
         DK-2800 Kgs.\ Lyngby, Denmark}

\author{Antti--Pekka Jauho}
\email{antti@mic.dtu.dk}

\affiliation{Mikroelektronik Centret,
         Technical University of Denmark,
         \O rsteds Plads, Bldg.\ 345 east,
         DK-2800 Kgs.\ Lyngby, Denmark}

\date{\today}

\begin{abstract}
We present a quantum theory of the shuttle instability in
electronic transport through a nanostructure with a mechanical
degree of freedom. A phase space formulation in terms of the
Wigner function allows us to identify a cross-over from the
tunnelling to the shuttling regime, thus extending the previously
found classical results to the quantum domain. Further, a new
dynamical regime is discovered, where the shuttling is driven
exclusively by the quantum noise.
\end{abstract}

\pacs{05.60.Gg, 73.23.Hk, 85.85.+j}

\maketitle

Advances in microfabrication technology are pushing today's
microelectromechanical systems (MEMS) towards the nanometer
regime, and the emerging new technology of nanoelectromechanical
systems (NEMS) is expected to play an important role in the
future. The ubiquitous quantum mechanical effects affecting the
performance of these devices present many theoretical challenges,
only few of which have so far been addressed in the literature.
The purpose of this Letter is to present a fully quantum theory
for an electromechanical instability in a generic NEMS device, the
single-electron shuttle first studied by Gorelik et al
\cite{gor-prl-98}.

This device consists of a movable single-electron transistor (SET)
and exhibits an electromechanical instability from the standard
tunnelling regime to a new regime in which the SET oscillates and
carries an integer number of electrons per a cycle (shuttle
regime). Since the original suggestion \cite{gor-prl-98}, there
has been increasing interest in the shuttle phenomenon
\cite{isa-phb-98,nis-prb-01,fed-epl-02,nor-prb-02,arm-prb-02,nis-prl-02,smi-prb-02,gor-nat-01,fed-preprint-02}:
e.g.\ by incorporating the shot noise due to the electron transfer
\cite{isa-phb-98}, gate effects \cite{nis-prb-01}, the coherence
effects in the electronic subsystem \cite{fed-epl-02} and strong
dissipation of the oscillator energy \cite{nor-prb-02}. Only very
recently the quantum mechanical treatment of the oscillations in
various modifications of the shuttle setup has been considered
\cite{arm-prb-02,smi-prb-02}.

The classical theory of shuttle transport has been used
\cite{nis-prb-01,fed-epl-02} to describe the experiments on ${\rm
C}_{60}$ single-electron transistor \cite{par-nat-00}, where the
oscillations of the center of mass of the molecule were found to
be important. However, also explanations based on incoherent
phonon assisted tunnelling theory which do not take into account
the correlation between the coherent oscillator motion and the
electron transfer seem to yield reasonable predictions for the
I--V curves \cite{boe-epl-01,mcc-preprint-02}.

Therefore, a fully coherent quantum mechanical treatment of the
oscillator which does take into account possible correlations and
which would, in principle, allow to join the two approaches into a
unified framework is a most desirable theoretical task. In this
study we present an attempt for such a treatment for a simplest
model supporting the shuttling transition in the classical regime.
A complementary study to ours already exists
\cite{fed-preprint-02} which uses a similar model, albeit without
any mechanical damping. However, the quasiclassical expansion of
the tunneling term used in that work does not give access to the
purely quantum phenomena discussed below.

We demonstrate that the characteristic strong correlation between
the oscillator motion and the electron transfer persists even in
the quantum regime. The noise generated by various sources (shot
and thermal, both having quantum components) is found to be very
important for the phenomenon. Not only does it smear the classical
transition found in \cite{gor-prl-98} into a crossover with a
considerably shifted position in parameter space compared to the
classical ``mean field" study \cite{gor-prl-98,fed-epl-02} but the
quantum component of the shot noise also generates the shuttle
instability even in the classically stable region with zero
electric field.

Using the generalized master equation approach suggested in
\cite{arm-prb-02} we study a simple model motivated by several
previous studies
\cite{gor-prl-98,fed-epl-02,mcc-preprint-02,fed-preprint-02}.
Namely, we consider an oscillating nanoscopic grain with only one
electronic level (strict Coulomb blockade regime) coupled to two
leads. The oscillator degree of freedom is treated fully quantum
mechanically in our approach. We also account for the oscillator
damping, the thermal noise due to the oscillator-bath coupling,
and the shot noise due to the quantized electron transfer.

The Hamiltonian of the model reads
\begin{equation}
\begin{split}
    H = H_{\rm osc} + \eps_0 c^{\dag}_0 c_0 +
    \sum_{k;\alpha=L,R}(\eps_{k\alpha}-\mu_{\alpha})c_{k\alpha}^{\dag}c_{k\alpha}+
    H_B \\
    - eE x c^{\dag}_0 c_0 + \sum_{k;\alpha=L,R} (T_{k\alpha0}(x) c_{k\alpha}^{\dag} c_0 +
    h.c.) + H_{\rm int}
\end{split}
\end{equation}
where the first three terms describe the free evolution of a
linear harmonic oscillator $H_{\rm
osc}=\tfrac{p^2}{2m}+\tfrac{m\om^2x^2}{2}$, the single electronic
level on the grain, and the two noninteracting leads symmetrically
biased by voltage $-V$ (i.e.\
$\mu_L=\tfrac{eV}{2},\,\mu_R=-\tfrac{eV}{2}$ with $e, V> 0$; all
electronic energies are measured with respect to the equilibrium
chemical potential of the leads), respectively. The term $H_B$
describes a generic Ohmic heat bath \cite{weiss}. The first
interaction term accounts for the electrostatic coupling between
the position of the grain and the occupation of the grain
electronic level; $E$ is the electric field caused by the bias
applied between the leads and/or gate voltage
\cite{nis-prb-01,mcc-preprint-02}. This electric field causes a
shift of the charged oscillator equilibrium by
$d=\tfrac{eE}{m\om^2}$. The second term is the coupling of the
leads to the single electronic state on the grain with
oscillator-dependent hopping amplitudes $T_{kL0}(x) = t_L
\exp(-\tfrac{x}{\lambda}),\, T_{kR0}(x) = t_R
\exp(\tfrac{x}{\lambda})$ where $\lambda$ is the electron
tunnelling length. The last term $H_{\rm int}$ describes the
coupling of the heat bath to the oscillator which is linear in the
oscillator coordinate \cite{weiss}. The net effect of this
coupling is the oscillator mechanical damping characterized by a
constant $\gamma$ and the Langevin random force that the bath
exerts on the oscillator and which depends on the temperature of
the bath $T$.

By projecting out the leads and the thermal bath we arrive at a
Markovian generalized master equation (GME) for the density matrix
$\rho(t)$ of the system composed of the grain and the harmonic
oscillator. Since the electronic transfer rate in the shuttling
regime is comparable with the frequency of the oscillations we
employed the singular coupling limit \cite{spo-rmp-80,kim-pra-01}
appropriate for rapidly decaying systems. This approximation is
valid for bias much larger than any energy scale of the system,
i.e.\ $eV \gg \hbar\om ,\eps_0$. Moreover, we also assume $eV \gg
k_B T$ which is reasonable in the present physical context. The
mechanical damping due to the heat bath is on the other hand
treated within the standard weak coupling theory
\cite{koh-jcp-97}.

Our GME reads
\begin{align}
   \dot{\rho}(t) &= \mathcal{L}\rho(t) = (\mathcal{L}_{\rm coh} + \mathcal{L}_{\rm driv}
   + \mathcal{L}_{\rm damp})\rho(t) \label{supermatrix}\\
\intertext{with}
   \mathcal{L}_{\rm coh}\rho &= \frac{1}{i\hbar}[H_{\rm osc}
   + \eps_0 c^{\dag}_0 c_0- eE x c^{\dag}_0 c_0,\rho] \ ,\\
   \mathcal{L}_{\rm driv}\rho &= -\frac{\Gamma_L}{2}
   \bigl(c_0 c^{\dag}_0 e^{-\frac{2x}{\lambda}}\rho -2 c^{\dag}_0 e^{-\frac{x}{\lambda}}
   \rho e^{-\frac{x}{\lambda}} c_0 + \rho e^{-\frac{2x}{\lambda}}c_0
   c^{\dag}_0\bigr)  \notag \\
   &-\frac{\Gamma_R}{2}
   \bigl(c^{\dag}_0 c_0 e^{\frac{2x}{\lambda}}\rho -2 c_0 e^{\frac{x}{\lambda}}
   \rho e^{\frac{x}{\lambda}}c^{\dag}_0 + \rho e^{\frac{2x}{\lambda}}c^{\dag}_0
   c_0\bigr) \ , \\
   \mathcal{L}_{\rm damp}\rho &= -\frac{i\gamma}{2\hbar}[x,\{p,\rho\}] -
   \frac{\gamma m\omega}{\hbar}(\bar{N}+1/2)[x,[x,\rho]] \ .\label{TI}
\end{align}
Here, $[,]$ and $\{,\}$ denote the commutator and the
anticommutator, respectively and
$\bar{N}=(\exp(\tfrac{\hbar\om}{k_B T})-1)^{-1}$ is the mean
oscillator occupation at the bath temperature.

The first term in Eq.\ \eqref{supermatrix} is the free coherent
evolution of the oscillator and the grain level while the second
one describes the transfer of electrons via the
oscillator-position-dependent tunnel junctions from the left
reservoir to the grain level (the term with $\Gamma_L$) and from
the level to the right reservoir (the term with $\Gamma_R$). The
opposite processes can be neglected due to the assumption $eV \gg
\hbar\om ,\eps_0, k_B T$. The transfer rates equal $\Gamma_{L,R}=
\tfrac{2\pi}{\hbar}|t_{L,R}|^2\mathcal{D}_{L,R}$ with the constant
densities of states of the leads $\mathcal{D}_{L,R}$. The third
term accounts for the interaction of the oscillator with the heat
bath. Translational invariance of the damping, positivity of the
density matrix, and relaxation towards canonical equilibrium
cannot be achieved simultaneously with any Markovian damping
kernel and one has to sacrifice at least one of these properties.
The most physical choice is to relax the positivity
\cite{koh-jcp-97}. We checked the magnitude of breaking of the
positivity in our calculations and found it irrelevant. Moreover,
it only occurs for large values of $\gamma$ out of the shuttling
regime.

It can be shown that the electronic off-diagonal elements of the
density matrix are decoupled from the diagonal ones and, moreover,
decay to zero in the stationary state. Therefore, it is sufficient
to consider only the electronic diagonal elements:
$\rho_{00}(t)=\langle 0\,|\rho(t)|0\rangle$ and
$\rho_{11}(t)=\langle 1\,|\rho(t)|1\rangle$, where
$|1\rangle=c_0^{\dag}|0\rangle$. These objects are still full
density matrices in the oscillator space and satisfy
\begin{equation}\label{GME}
\begin{split}
    \dot{\rho}_{00}(t) &= \frac{1}{i\hbar} [H_{\rm osc},\rho_{00}(t)]
    - \frac{\Gamma_L}{2}(e^{-\frac{2x}{\lambda}}\rho_{00}(t)
    + \rho_{00}(t)e^{-\frac{2x}{\lambda}}) \\
    &+ \Gamma_R e^{\frac{x}{\lambda}}\rho_{11}(t)e^{\frac{x}{\lambda}}
    + \mathcal{L}_{\rm damp}\,\rho_{00}(t)\ , \\
    \dot{\rho}_{11}(t) &= \frac{1}{i\hbar}[H_{\rm osc}-eEx,\rho_{11}(t)]
    + \Gamma_L
    e^{-\frac{x}{\lambda}}\rho_{00}(t)e^{-\frac{x}{\lambda}}\\
    &- \frac{\Gamma_R}{2}(e^{\frac{2x}{\lambda}}\rho_{11}(t)
    + \rho_{11}(t)e^{\frac{2x}{\lambda}})
    + \mathcal{L}_{\rm damp}\,\rho_{11}(t).
\end{split}
\end{equation}
From the continuity equation for the electronic charge we may
deduce the following formula for the stationary current through
the grain (flowing from the left to the right lead):
\begin{equation}\label{current}
    I^{\rm stat}=e \Gamma_L{\rm Tr_{osc}}(e^{-\frac{2x}{\lambda}}\rho^{\rm
    stat}_{00})
    = e \Gamma_R{\rm Tr_{osc}}(e^{\frac{2x}{\lambda}}\rho^{\rm
    stat}_{11}) \ .
\end{equation}
The trace is carried out over the oscillator basis and $\rho^{\rm
stat}_{nn} = \lim_{t\to\infty}\rho_{nn}(t)$ .

We solved numerically the stationary version of the above
equations \eqref{GME}: $0=\sum_{l=1}^{2N^2}L_{kl}\rho^{\rm
stat}_l$ where the column vector $\rho^{\rm stat}_l$ consists of
the matrix elements of $\rho^{\rm stat}_{00},\rho^{\rm stat}_{11}$
and the (super-)matrix $L_{kl}$ of the dimension $2N^2\times 2N^2$
contains the appropriate coefficients of the linear system
\eqref{GME}. The density matrix was represented in the harmonic
oscillator basis which was truncated by taking up to $N=100$
lowest states which yields satisfactory numerical convergence in
our parameters range. We determined the unique (for
$\Gamma_R=\Gamma_L=\Gamma\neq 0$) null vector of the supermatrix
$L$ by using the Arnoldi iteration \cite{golub}.

\begin{widetext}

\begin{figure}
  \includegraphics[width=126mm]{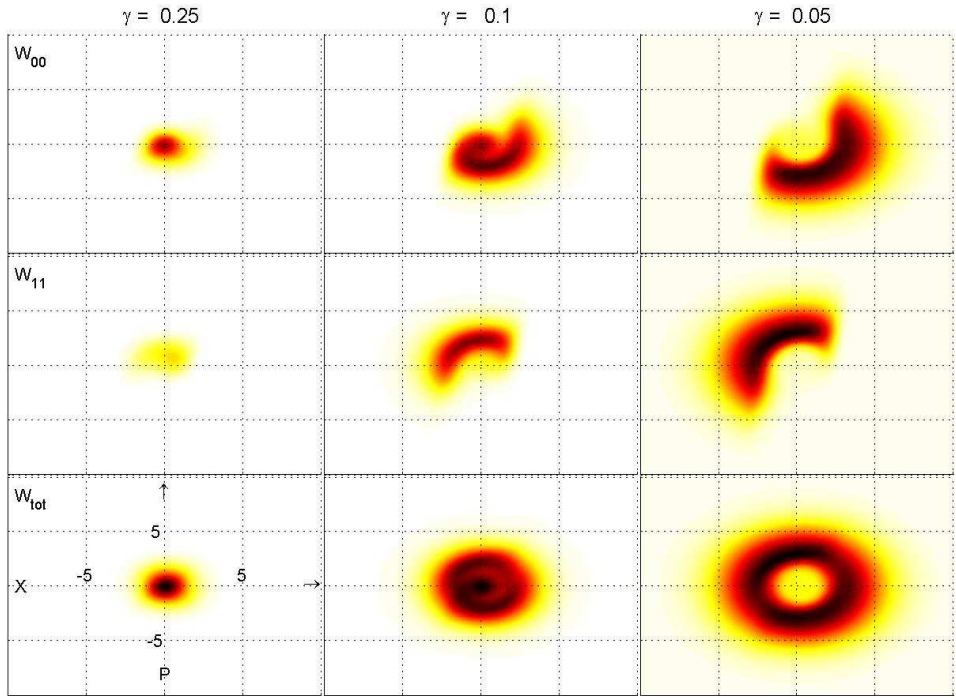}
  \caption{Phase space picture of the tunnelling-to-shuttling crossover.
  The respective rows show the Wigner distribution functions
  for the discharged ($W_{00}$), charged ($W_{11}$), and both ($W_{\rm tot}$)
  states of the oscillator in the phase space (horizontal axis --
  coordinate in units of $x_0=\sqrt{\hbar/m\om}$,  vertical axis -- momentum in $\hbar/x_0$).
  The values of the parameters are: $\lambda=x_0,T=0,d=0.5x_0,\Gamma=0.05\hbar\om$.
  The values of $\gamma$ are in units of $\hbar\om$. The Wigner functions are
  normalized within each column.}
  \label{fig1}
\end{figure}
\end{widetext}

Our approach differs from the one used in \cite{arm-prb-02} where
the explicit time integration scheme of the time-dependent
equation analogous to \eqref{GME} was used to determine the
stationary density matrix. We applied our method to the model of
\cite{arm-prb-02} and recovered the I-V curve results reported
there. Using the phase space analysis detailed below we fully
confirmed the shuttling interpretation based on the indirect
evidence from changes of the I-V curves with changing parameters.

The I-V curve (or other dependencies of the current on some
parameter) alone yields only an indirect evidence of shuttling and
may actually not be decisive whether the system is shuttling or
not (see e.g.\ \cite{boe-epl-01} versus \cite{fed-epl-02}).
Therefore, it is preferable to consider quantities which depend
also on the {\it state of the oscillator}. An excellent
visualization tool \cite{koh-jcp-97} for the description of the
joint electronic and oscillator properties are the Wigner
functions ($n=0,1$)
\begin{equation}
    W_{nn}(X,P) =
    \int_{-\infty}^{\infty}\frac{dy}{2\pi\hbar}\,\bigl\langle
    X-\frac{y}{2}\,|\rho_{nn}^{\rm stat}|X+\frac{y}{2}\bigr\rangle\,
    \exp\bigl(i\frac{Py}{\hbar}\bigr)
\end{equation}
yielding the charge-resolved quasiprobability distributions of the
oscillator in the phase space. These functions provide us with a
clear evidence of the transition from the incoherent tunnelling to
the coherent quasiclassical shuttling behavior with decreasing
damping coefficient.

Our focus is in the quantum effects on the shuttling transition
and, in particular, whether there is any transition in the quantum
regime at all. Therefore, we work in the strictly quantum regime
where the tunnelling length is comparable to the zero uncertainty
of the oscillator, $\lambda\sim x_0=\sqrt{\hbar/m\om}$, and where
only a relatively small number of oscillator states is excited.

In Fig.\ \ref{fig1} we depict the Wigner functions
$W_{00},W_{11}$, and $W_{\rm tot}=W_{00}+W_{11}$  showing the
crossover from the tunnelling to the shuttling regime with
decreasing damping. In the tunnelling regime (large $\gamma$) the
oscillator is located around the origin (or shifted origin when
charged) with no particular correlation between its charge state
and momentum (Wigner functions are centered around the origin with
some ``quantum fuzziness"). This is consistent with the quantum
incoherent tunnelling picture. On the other hand, in the shuttling
regime (small $\gamma$) the oscillator orbits almost classically
(ring-like shape of $W_{\rm tot}$ with a hole around the origin),
and shuttles the charge on its way from the left to the right lead
and returns empty back (half-moon shapes of $W_{00},W_{11}$). The
correlation between the charge state and the mechanical motion is
very strong. In the crossover region (medium $\gamma$) we can see
that both regimes of transport are contributing additively
(ring-like shape plus an incoherent peak around the origin of
$W_{\rm tot}$). The classical ``mean field" sharp transition of
\cite{gor-prl-98} between the tunnelling and the shuttling regime
is smeared into a crossover due to the noise. Also the position of
the crossover is substantially shifted with respect to the
classical values ($\gamma_{\rm cross}$ is $\sim 5$ times larger
than the classical value) which is attributed to the deeply
quantum (and therefore noisy) regime. The classical picture is
expected to emerge in the quasiclassical limit $d,\lambda \gg
x_0$.

\begin{figure}
  \includegraphics[width=80mm]{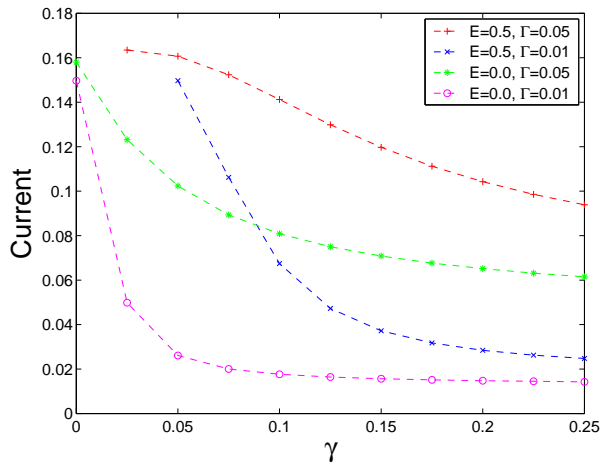}
  \caption{$I-\gamma$ curve. The $\gamma$-dependence of the stationary
current through the grain for different transfer rates and
electric fields. Their values are $d=0.5x_0,\Gamma=0.05\hbar\om$
(pluses; corresponds to Fig.~\ref{fig1}),
$d=0.5x_0,\Gamma=0.01\hbar\om$ (circles),
$d=0.0,\Gamma=0.05\hbar\om$ (asterisks), $d=0.0,\Gamma=0.01\hbar$
(crosses). Other parameters are $\lambda=x_0,T=0$. The current is
in units of $e\om$ while $\gamma$ in $\hbar\om$.}\label{fig2}
\end{figure}

In Fig. \ref{fig2} we plot the $\gamma$-dependence of the
stationary current through the grain for different transfer rates
$\Gamma$ and electric fields (measured in $d$). For the two curves
with $d \neq 0$ we can see the rise of the current in the
crossover region from the tunnelling-limited values proportional
to $\Gamma$ to shuttling-mediated quantized value of one shuttled
electron per each cycle ($1/2\pi\approx 0.16$ in our units;
independent of $\Gamma$) in agreement with the classical results.
More surprisingly, also the results for the case of zero electric
field $d=0$ show clear signs of shuttling crossover in the
$I-\gamma$ curve for small enough mechanical damping. Classically
there is no shuttling transition for $d=0$ regardless of the other
parameters values \cite{fed-epl-02} and the $I-\gamma$ curve is
constant.  Therefore, this shuttling must be driven purely by the
quantum component of the shot noise (proportional to $\hbar$).
Also the phase space pictures of this regime reveal onset of
shuttling transport.

Finally, we comment on the effect of the temperature. For a
nonzero temperature the shuttling transition within our model is
facilitated by the increase of the mechanical noise driving the
transition. The deteriorating effect of the temperature on the
transition is not included in the model due to the high bias
assumption. The development of the theory for a finite bias is
under way.

To summarize, we have presented a quantum theory of the shuttling
transition in fully developed Coulomb blockade regime. Using the
Wigner functions as a phase space visualization method we have
exhibited a clear crossover from the tunnelling to the shuttling
regime of the transport as a function of the mechanical damping
parameter. The effect of noise on the transition in the deeply
quantum regime ($\lambda\sim x_0=\sqrt{\hbar/m\om}$) is pronounced
and can even trigger the transition in the classically stable
regime with the zero electric field.

The authors want to thank the members of the Chalmers group,
A.~Wacker, and B.~Velick\'y for helpful discussions. Advice from
T.~Eirola concerning numerical methods was indispensable. Support
of the grant 202/01/D099 of the Czech grant agency for one of us
(T.N.) is also gratefully acknowledged.


\end{document}